\documentclass[traditabstract,rnote]{aa} % for the research note
\usepackage{graphicx}
\usepackage{txfonts}
\usepackage{amssymb}
\usepackage{natbib}
\usepackage{algorithmic}
\def\eq#1{\begin{equation} #1 \end{equation}}
\begin{document}
   \title{Constraints on the height of the inner disk rim in pre-main-sequence stars}

   %\subtitle{}

   \author{D. Vinkovi\'c
          \inst{1,2}
          }

   \institute{Physics Department, University of Split, Nikole Tesle 12, HR-21000 Split, Croatia\\
              \email{vinkovic@pmfst.hr}
         \and
             Science and Society Synergy Institute, Bana J. Jela\v{c}i\'{c}a 22B, HR-40000 \v{C}akovec, Croatia\\
               \email{dejan@iszd.hr}
             }

   \date{}

% \abstract{}{}{}{}{}
% 5 {} token are mandatory

  \abstract
   {The structure of inner region of protoplanetary disks around young pre-main-sequence stars is still poorly understood. This part of the disk is shaped by various forces that influence dust and gas dynamics, and by dust sublimation, which creates abrupt drops in the dust density. This region also emits strong near-infrared excess that cannot be explained by classical accretion disk models, which suggests the existence of some unusual dust distribution or disk shape. The most prevalent explanation to date is the puffed-up inner disk rim model, where the disk exhibits an optically thin cavity around the star up to the distance of dust sublimation. The critical parameter in this model is the inner disk rim height $z_{\rm max}$ relative to the rim distance from the star $R_{\rm in}$. Observations often require $z_{\rm max}/R_{\rm in}\gtrsim0.2$ to reproduce the near-infrared excess in the spectra. We compile a comprehensive list of processes that can shape the inner disk rim and combine them into a self-consistent model. Two of them, radiation pressure force and the gas velocity profile, have never been applied in this context before. The aim was to find the most plausible theoretical values of $z_{\rm max}/R_{\rm in}$. The results show that this value is $\lesssim$0.13 for Herbig Ae stars, $\lesssim$0.11 for T Tau stars, and $\lesssim$0.10 for young brown dwarfs. This is lower than the observational requirements for Herbig Ae stars. We argue that the same problem exists in T Tau stars as well. We conclude that the puffed-up inner rim model cannot be the sole explanation for the near-infrared excess in young pre-main-sequence stars.




   }


   \maketitle
%
%________________________________________________________________

\section{Introduction}

Protoplanetary disks surrounding young pre-main-sequence stars have been a subject of intense theoretical and observational scrutiny in recent decades \citep{Williams}. Since these disks are expected to host planet formation, particular focus has been on dust properties and dust distribution within the disks. While this led to an advanced understanding of the dust evolution beyond $\sim$1 AU from the star, the inner disk region within $\lesssim$1 AU remained a controversial topic \citep{Millan-Gabet,DullemondARAA}. The main obstacle is the inability of current telescopes to spatially resolve this region, except by the near-infrared interferometry, which in itself depends on modeling assumptions and currently provides valuable but still limiting information on the disk structure \citep{Wolf}.

The inner disk rim is observable at near-infrared wavelengths because dust emission in this part of the disk comes from dust temperatures above 1,000K. Even the most resilient dust grains sublimate in these conditions, but the exact sublimation distance from the star depends on the grain size and chemistry and on the local gas pressure. Removal of dust grains creates large steps in the opacity gradient, which complicates modeling the dusty disk structure \citep{KMD09,Vinkovic12}. Modeling efforts have therefore been devoted to approximate methods, motivated by the peculiar near-infrared bump in the spectral energy distribution of many pre-main-sequence stars.

This bump was first noticed in Herbig Ae/Be stars by \citet{Hillenbrand}, while \citet{Chiang} showed that ordinary accretion disk models do not produce enough flux to explain the bump. \citet{DDN} introduced the most popular explanation for this phenomenon to date, based on the assumption that the disk gas is optically thin close to the star in the zone where dust grains cannot survive. This allows stellar radiation to heat the whole vertical profile of the inner dusty disk rim, including the disk interior, which is typically much colder than the disk surface. This causes the disk to expand vertically, or "puff-up", which also increases the emitting area of the hot dusty disk rim. The near-infrared bump is then merely a measure of how much the disk expands vertically.

That assumption of an optically thin hole was confirmed by near-infrared interferometry \citep{Millan-Gabet01,Millan-Gabet}, which boosted the popularity of the puffed-up disk model. The model was additionally improved \citep[e.g.][]{DD04a,DD04b,Isella05,Tannirkulam07,Thi} and is still the most prevalent description of the inner disk structure \citep{DullemondARAA}. However, over the years, two main concerns about the plausibility of the model have been raised. One concern is that the near-infrared excess predicted by the puffed-up disk model is too low to explain the strongest near-infrared bumps, unless physically unrealistic vertical puffing is invoked, with heights several times above hydrostatic equilibrium \citep{DAlessio,VIJE,Meijer,Schegerer,Acke,Verhoeff10,Verhoeff11,Mulders,McClure2013a}. In addition, as \citet{Mulders} explain in their Appendix C, such strong puffing up casts a shadow over large parts of the outer disk, which would decrease the disk temperature and reduce its mid-infrared emission below the observed levels.
Moreover, the claim that an optically thin hole does not contribute to the near-infrared excess and images has been challenged recently by some interferometry observations \citep{Akeson2005,Tannirkulam08,Benisty10,Benisty11} and also theoretically
\citep{Vinkovic06,KMD09}.

The other concern is that the theoretical approaches to the modeling of the inner disk have been lacking some critical physical elements and therefore cannot correctly describe its structure. Small grains have often been included into the dust opacity of the inner rim, but now we know that the shape of inner disk rim is dictated solely by big grains ($\gtrsim 1\mu m$) because they are more resilient to sublimation than smaller grains. Hence, big grains populate the inner disk surface and shield smaller grains in the disk interior from direct stellar radiation \citep{KMD09,Vinkovic12}. An unexpected consequence of radiative transfer in big grains is that the location of the inner disk rim becomes a nontrivial problem, with the temperature inversion effect producing a local temperature maximum within the dust cloud (see the appendix). Here we focus only on the height and shape of optically thick disk, where we use the simple prescription for the rim radius presented by \citet{Vinkovic06}.

Moreover, processes such as vertical gas velocity profile \citep{TakeuchiLin02} or radiation pressure on dust grains \citep{TakeuchiLin03,Vinkovic09} have been lacking from the debate on the puffed-up rim model. In this paper we construct a model that takes into consideration a comprehensive list of physical processes that influence the rim shape. Our goal is to find $z_{\rm max}/R_{\rm in}$ for a wide range of disk parameters and determine constraints on the applicability of the puffed-up inner rim model.

\section{Physics of the inner disk structure}

   \begin{figure}
   \centering
   \includegraphics[width=\columnwidth]{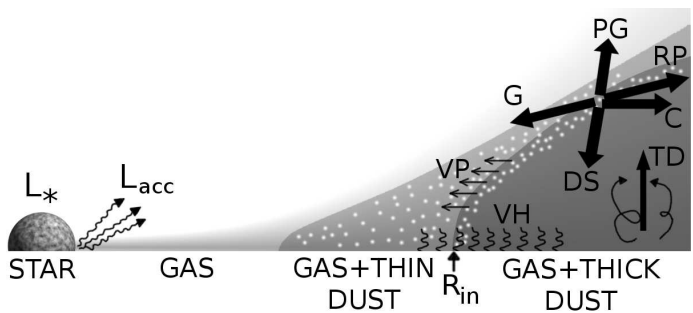}
      \caption{Sketch of the disk and processes used in our calculations. Three shades of gray show the regions clear of dust, with optically thin dust, and with optically thick dust. The beginning of the optically thick disk at $R_{\rm in}$ is defined as the disk rim. White dots mark the zone dominated by big grains, $L_*$ is the stellar luminosity, $L_{\rm acc}$ is the accretion luminosity, $VH$ the viscous heating, $TD$ the turbulent diffusion, $G$ the gravity, $C$ the centrifugal force, $DS$ the dust settling due to the gas drag force, $PG$ the gas pressure gradient, $RP$ the radiation pressure, and $VP$ is the gas velocity profile.
              }
         \label{FigSketch}
   \end{figure}

We implemented various physical processes discussed previously in the context of protoplanetary disks, although not necessarily applied to the problem of inner disk rim. A detailed list of equations used in our calculation is given in the appendix, while here we present only a short version. A list of the physical processes together with a sketch of the disk geometry is given in Figure \ref{FigSketch}. We considered an axially symmetric disk, described with a cylindrical coordinate system where $r$ is in the disk midplane and $z$ is in the direction of the disk symmetry axis. We used 1AU as the unit of distance and one year as the unit of time.

The concept of puffed-up inner rim model is related to the optically thick disk surface. Hence, we define the location of the inner rim $R_{\rm in}$ as the midplane distance of optically thick dust from the star \citep{Vinkovic06}
  \eq{\label{Rin_thin_thick}
   R_{\rm in}=0.0344\Psi\left( \frac{1500 K}{T_{\rm sub}} \right)^2 \sqrt{\frac{L_*+L_{\rm acc}}{L_\odot}} \,\,\,\,\,\mathrm{[AU]},
 }
where $L_*$ is the stellar luminosity, $L_{\rm acc}$ is the accretion luminosity, and $\Psi=2$ is the coefficient accounting for the effect of optically thick dust back-heating. This is also the place of local temperature maximum equal to local values of $T_{\rm sub}$ \citep[see Figure 3 in][]{KMD09}. High-resolution radiative transfer calculation shows that the entire inner rim experiences temperatures close to $T_{\rm sub}$, with dominantly radial temperature variation close to the rim \citep{Vinkovic12}. Hence, a vertically isothermal disk is a good approximation as long as we consider only the rim properties. The disk temperature in this narrow zone is then described with
 \eq{\label{temperature_change}
 T(r)=T_{\rm sub}\left(\frac{r}{R_{\rm in}}\right)^q ,
 }
where we treat $q$ as a free parameter, while dust sublimation $T_{\rm sub}$ depends on the local gas density $\rho_{\rm gas}$ \citep{Pollack}.

The gas is in vertical hydrostatic equilibrium
 \eq{\label{gas_density}
 \rho_{\rm gas}(r,z)  =  \rho_0 \left(\frac{r}{R_{\rm in}}\right)^p \exp\left(-\frac{z^2}{2H_p^2}\right) ,
 }
where $\rho_0 = \rho_{\rm gas}(R_{\rm in},0)$ is the midplane gas density at the inner disk rim, $H_p(r)=c_s(r)/\Omega_K(r)$ is the disk scale height that varies with $r$, $c_s$ is the isothermal sound speed, and $\Omega_K$ is the Keplerian angular velocity at the midplane.

Accretion of the gas onto the star contributes with accretion luminosity to the total energy that heats the inner rim \citep{Muzerolle}. To accommodate for this effect we calculated the accretion luminosity using the accretion rate $\dot{M}_{\rm acc}$ derived from the vertical profile of the gas density and the horizontal velocity of gas flow (see the appendix). We used the horizontal component of the gas velocity derived by \citet{TakeuchiLin02}
 \eq{
 \mathrm{v}_{\rm r,gas}(r,z)=-\frac{\alpha c_s H_p}{r}\left( 3p+2q+6+\frac{5q+9}{2}\frac{z^2}{H_p^2} \right)
                   \left[\mathrm{ \frac{AU}{year} }\right],
 }
who used the $\alpha$-prescription for the viscous effect of turbulence \citep{ShakuraSunyaev,Pringle}.
The inward accretion exists only if
 $
 p+3(q+1)/2 > -2
 $ \citep{TakeuchiLin02}.
The accretion rate depends on the distance from the star, except for $q=-0.5$ when the radial velocity is independent of $r$ and the disk is in steady-state. We focused our attention on a very small part of the disk around $R_{\rm in}$, where $\dot{M}_{\rm acc}$ varies slowly with $r$. Hence, we define the accretion rate as $\dot{M}_{\rm acc}(R_{\rm in})$. In the appendix, we show $\dot{M}_{\rm acc}\propto \rho_0$ and how we derive $\rho_0$ from $\dot{M}_{\rm acc}$, which is in turn derived from $L_{\rm acc}\propto \dot{M}_{\rm acc}$.

The local viscous dissipation of energy in accretion disks contributes to the local disk temperature, and we determined this effect. Hence, the net temperature at $R_{\rm in}$ is the combination $T_{\rm sub}^4=T_{\rm irr}^4+T_{\rm visc}^4$ \citep{CieslaCuzzi} of the temperature from stellar irradiation and from viscous heating (see the appendix for details). This condition has to be checked iteratively during the determination of $R_{\rm in}$.

Another process we considered is the radial migration of dust particles in the gas disk. We followed the prescription by \citet{TakeuchiLin02}, where dust particles are driven by gas drag and their radial velocity depends on the distance from the midplane. The source of this vertical velocity variation is the gas pressure gradient, which causes two effects: gas rotation is different from Keplerian rotation above the midplane, and the gas drag force depends on gas density.

Without gas drag, dust particles would follow Keplerian orbits. We considered particles smaller than the mean free path of gas molecules. Their motion is described with Epstein's gas drag law. We also considered turbulent disks where the gas turbulent motion stirs up dust particles to higher altitudes and prevents dust sedimentation. The equilibrium distribution of dust particles is reached when the sedimentation due to gas drag and the diffusion due to turbulence are balanced. Using the $\alpha$-prescription for the viscous effect of turbulence, the dust density distribution is \citep{TakeuchiLin02}
 \eq{
 \rho_{\rm dust}(r,z)=\rho_d(r) \exp\left( -\frac{z^2}{2H_p^2} - \frac{S_c\,\Omega_K\,\, t_0}{\alpha}
                  \left(\exp\frac{z^2}{2H_p^2}-1\right) \right),
 }
where $\rho_d(r)$ is the midplane dust density profile and $t_0 (r)$ is given in the appendix. The Schmidt number $S_c$ is a measure of the coupling between the particles and the gas. Even though $S_c\sim 1$ for the particle sizes used in our paper, we adopted the expression for $S_c$ suggested by \citet{Youdin} (see the appendix for details).

\citet{TakeuchiLin03} showed that stellar radiation pressure force slowly erodes the dusty disk surface, but they did not investigate the impact of this force on the surface of inner disk rim. We used their methodology to include the radiation pressure force into the dust velocity profiles. We ignored the complicated infrared radiation pressure force of photons from the disk interior because this force can only enhance the disk erosion \citep{Vinkovic09}. Hence, the radial component of dust velocity is \citep{TakeuchiLin03}
 \eq{
 \mathrm{v}_{\rm r,dust}(r,z)=\frac{\mathrm{v}_{\rm r,gas}+(\beta-\eta)\,r\,t_{\rm drag}\,\Omega_K^2}
                               {1+t_{\rm drag}^2\Omega_K^2} \left[\mathrm{ \frac{AU}{year} }\right] ,
 }
where $\beta$ is the ratio between the stellar radiation pressure force and gravity and $\eta$ is a parameter given in the appendix.

Our goal is to find the relative height $z_{\rm max}/R_{\rm in}$ above which the outward flow \hbox{$\mathrm{v}_{\rm r,dust}>0$} removes dust from the disk rim. The disk volume below the height $z_{\rm max}$ can be populated entirely by dust on long timescales, but above this limit dust grains are quickly eroded. From the radiative transfer point of view, the ratio $z_{\rm max}/R_{\rm in}$ measures the maximum fraction of the stellar radiation intercepted by the disk rim, which in turn is the maximum energy reemitted into the near-infrared.

\begin{table}
\caption{Parameters of our disk model and their range of values}
\label{parameters}
\centering
\begin{tabular}{l c c c}
\hline\hline
\noalign{\smallskip}
 & Herbig Ae & T Tau & brown dwarf \\
$L_* [L_\odot]$ & 40 & 0.5 & 0.004 \\
$M_* [M_\odot]$ & 2.5 & 0.5 & 0.07 \\
$T_* [K]$ & 10,000 & 4,000 & 3,000 \\
$L_{\rm acc} [L_\odot]$ & 0.5-8.5 & 0.02-0.2 & 0.0005-0.005 \\
\hline
\noalign{\smallskip}
$a\, [\mu m]$ & \multicolumn{3}{c}{1, 10, 100} \\
$\rho_{\rm grain}\, [\mathrm{g/cm^3}]$ & \multicolumn{3}{c}{0.1$^\dag$, 3} \\
$\alpha$ & \multicolumn{3}{c}{0.001, 0.01, 0.1} \\
$p$ & \multicolumn{3}{c}{0, -0.5, -1, -1.5, -2, -2.5, -3} \\
$q$ & \multicolumn{3}{c}{0, -0.5, -1, -1.5, -2} \\
\hline
\noalign{\smallskip}
\multicolumn{4}{l}{$^\dag$ A highly porus, fluffy grain aggregate.} \\
\end{tabular}
\end{table}

\section{Results}

   \begin{figure}
   \centering
   \includegraphics[bb=5 0 390 295, width=\columnwidth]{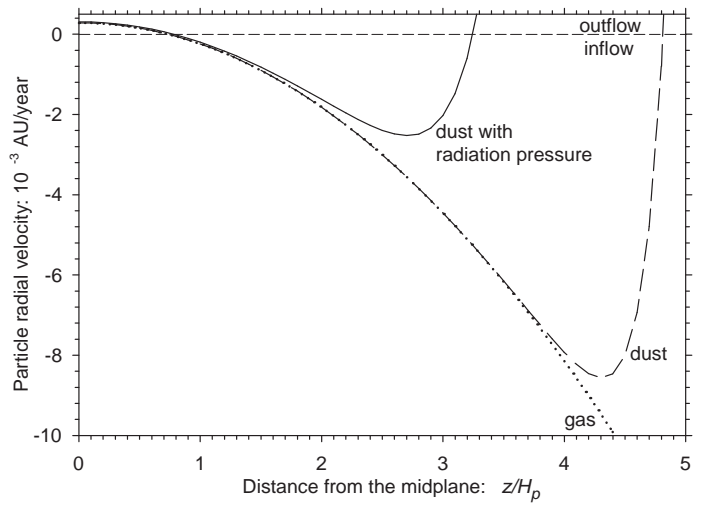}
      \caption{Example of model results showing the variation of horizonal velocities with height above the midplane (measured in scale heights). Lines show velocities of gas (dotted line), dust without the force of radiation pressure (long-dashed line), dust with the force (solid line), and zero velocity separating inflow from outflow (horizonal short-dashed line). Model parameters used in this result are $T_*$=10,000K, $L_*=40L_\odot$, $L_{\rm acc}=5L_\odot$, $M_*=2.5M_\odot$, $q=-0.5$, $p=-2.25$, $a=1\mu m$, $\rho_{\rm grain}=$3 g/cm$^3$, and $\alpha=0.01$. The model output is $\rho_0=2.6\times 10^{-9}$g/cm$^3$, $\dot{M}_{\rm acc}=1.7\times 10^{-7}M_\odot/$year, and $R_{\rm in}=0.54$ AU. The dusty disk is truncated at a height of $z_{\rm max}=3.24 H_p = 0.11R_{\rm in}$ (the point where dust turns from inflow to outflow).
              }\label{FigVelocity}
   \end{figure}
   \begin{figure}
   \centering
   \includegraphics[bb= 5 0 420 195, width=\columnwidth]{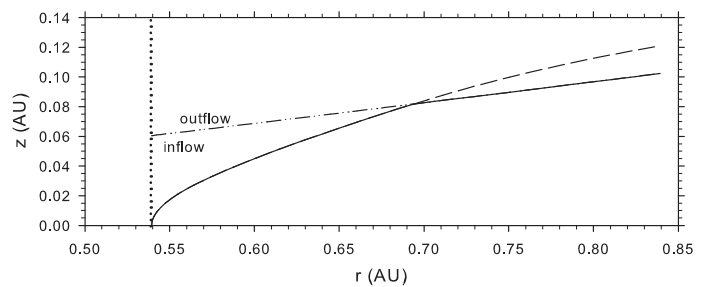}
      \caption{Shape of inner disk surface defined as the radial optical depth of $\tau=1$ (solid line). The model parameters are the same as in Figure \ref{FigVelocity}. The inner disk radius $R_{\rm in}$ is marked with the dotted line (the velocity profiles in Figure \ref{FigVelocity} follow this line). The dust density scale is set by using the spatial length of 10$^{-5}$AU in the midplane between $R_{\rm in}$ and $\tau=1$. The radiation pressure force dominates the region above the dot-dot-dashed line and moves dust particles outward. The disk surface without radiation pressure would extend as far as the dashed line.}\label{FigProfile}
   \end{figure}
   \begin{figure}
   \centering
   \includegraphics[width=\columnwidth]{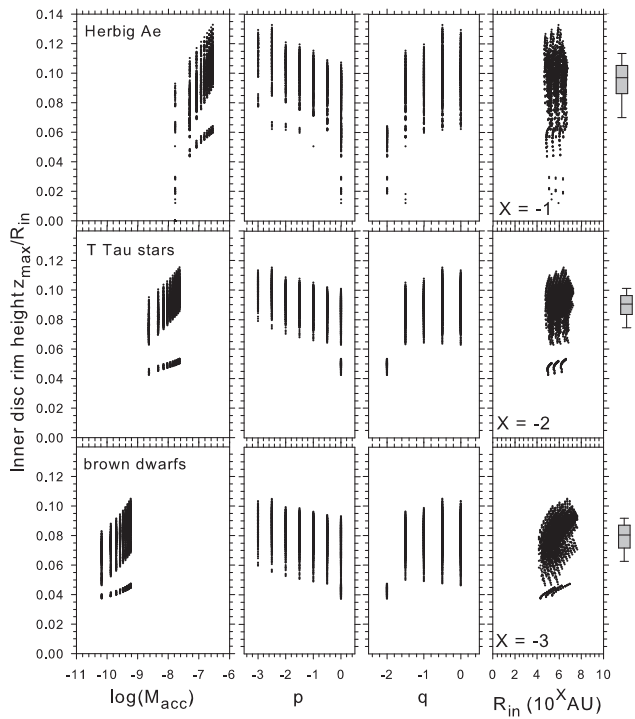}
      \caption{Distribution of $z_{\rm max}/R_{\rm in}$ values relative to the mass accretion rate $\dot{M}_{\rm acc}$ (in units of $M_\odot yr^{-1}$), the power law $p$ of the disk density change from equation \ref{gas_density}, the temperature power law $q$ from equation \ref{temperature_change}, and the inner disk radius $R_{\rm in}$ from equation \ref{Rin_thin_thick}.
      Results are shown for three types of objects, as indicated in the figure: Herbig Ae, T Tau, and brown dwarfs.
      The box-and-whisker plots at the far right show the range of results for each object: boxes show the median line and the first and third quartile, while the whiskers show the 10$^{\rm th}$ and 90$^{\rm th}$ percentile.
      }\label{FigZmax}
   \end{figure}

We explored a wide range of parameters (see Table \ref{parameters}) for Herbig Ae and T Tau stars and brown dwarfs. The ranges of accretion luminosities are based on observational data \citep{Randich,Herczeg,Donehew,Mendigutia}. In total, more than 6,000 models were calculated. Some models result in unrealistically large disk masses (that would lead to gravitational instabilities), but nevertheless, we did not exclude them from the analysis. The accretion rates are lower than 10$^{-6}$M$_\odot yr^{-1}$, which is consistent with the assumption of optically thin gas within $R_{\rm in}$ \citep{Muzerolle}. The dust optical constants are taken from \citet{Dorschner}. We used two extreme types of grains: solids with 3g/cm$^3$ in density and highly porus, fluffy aggregates with 0.1g/cm$^3$.

Figure \ref{FigVelocity} shows a typical model outcome. The horizontal component of the dust velocity at $R_{\rm in}$ depends on height and directs the flow of gas. The flow varies with height between outflow and inflow regimes. Dust motion is dictated by gas within a couple of gas scale heights from the midplane. Depending on the radial gas density profile, this flow can start as a slow outflow in the midplane before it turns into inflow above the midplane. The gas flow would dictate the dust dynamics even at larger heights, but the radiation pressure force starts to dominate and dust drifts into outflow. The transition from inflow to outflow is at $z_{\rm max}$. Without the radiation pressure dust grains would follow the gas until gas and dust decouple and the results would be identical to the work by \citet{TakeuchiLin02}.

The inability of the dust to move inward in the zone above $z_{\rm max}$ has consequences for the shape of the inner disk surface. An estimate of the inner disk edge shape is shown in Figure \ref{FigProfile}, based on the model from Figure \ref{FigVelocity}. The shape is defined as the surface where the radial optical depth is $\tau=1$. The dust density scale $\rho_0$ is defined by setting the spatial depth of $\tau=1$ at $R_{\rm in}$ in the midplane (this depth is so low that $R_{\rm in}$ can be treated as the distance of optically thick surface from the star). The figure shows how the surface curves because of dust density drop with height \citep[as was also shown by][]{Tannirkulam07}, but only at heights below the dust outflow zone created by the radiation pressure force. The dust density can be increased to boost the disk surface height, but only within the dust inflow zone.

Dust sublimation does not enter as a sharp boundary condition into the shaping of the disk surface, as is typically assumed \citep[e.g.][]{Isella05,Tannirkulam07}, because big dust grains can survive in the optically thin zone closer to the star than $R_{\rm in}$ (see examples by \citet{KMD09} and \citet{Vinkovic12}). Hence, in this estimate we ignored optically thin dust within $R_{\rm in}$ because we have no information on its density structure \citep{Vinkovic12}.

The main result is shown in Figure \ref{FigZmax}. The values of $z_{\rm max}/R_{\rm in}$ in all models are $\lesssim$0.13 for Herbig Ae stars, $\lesssim$0.11 for T Tau stars, and $\lesssim$0.10 for young brown dwarfs. The figure shows that the values vary with $\dot{M}_{\rm acc}$, $p$, $q$ and $R_{\rm in}$, where variations of values used for $L_{\rm acc}$ create a spread in $\dot{M}_{\rm acc}$ and $R_{\rm in}$.

\section{Discussion}

If the near-infrared excess is purely caused by the dusty disk emission, the amount of near-infrared excess at ${\sim}2\mu m$ is a direct measure of the inner disk rim height \citep{VIJE,Mulders}. Measurements of spectral energy distribution of Herbig Ae stars imply that for many objects the rim height must be $z_{\rm max}/R_{\rm in}\gtrsim0.2$ \citep{Akeson2005,VIJE,Eisner2007,Meijer,Schegerer,Acke,Mulders}.
This implies that the inner rim heights shown in Figure \ref{FigZmax} are not sufficient to explain strong near-infrared bumps. Some additional sources of the near-infrared excess must be present.

Moreover, the heights in our model do not produce a self-shadowing of the disk (Figure \ref{FigProfile}), which was one of the important consequences of the puffed-up rim model \citep{DullemondARAA}. High values of rim heights were derived by \citet{DD04a} using hydrostatic equilibrium disk models with very steep radial density profiles. They demonstrated self-shadowing for $p=-4$, but this implies a disk with almost all of its mass within the inner 1 AU from the star, which is unrealistic \citep[as was pointed out by][]{DAlessio}. Their models did not take into account the role of big dust grains in shaping the rim either. Their improved models that combine hydrostatic equilibrium, vertical dust settling, and turbulent mixing \citep{DD04b} did not show the near-infrared bump. A comparison of these models with the data was performed recently by \citet{Mulders}, who concluded that the models cannot explain the near-infrared bump.

Another important problem is the shape of the inner disk rim. The initial simple puffed-up rim model had a vertical wall \citep{DDN}, which does not agree with the near-infrared interferometry data \citep{DullemondARAA}. Therefore, the rim curvature has been modeled by several authors, but so far without the radiation pressure force and with a limited set of physical phenomena. Radiative transfer methods used by \citet{Isella05} and \citet{Tannirkulam07} (for Herbig Ae stars) and \citet{Mayne2010} (for brown dwarfs) were not appropriate for calculating temperature inversion effect at the inner rim. Furthermore, since they did not use grain size separation, where big grains shield small grains from the direct stellar radiation, the rim shape is not correct either. The drawbacks of these models were discussed by \citet{KMD09}, who also emphasized the importance of the temperature inversion effect in shaping the inner disk.

Our results imply that at least a fraction of the near-infrared excess must originate from some extension to the inner disk rim geometry. In the absence of detailed constraints, simple halo structures above the inner disk have been gaining popularity in recent years as a method to model the near-infrared observations \citep[e.g.][]{VIJE,Schegerer,Mulders2010,Verhoeff11,Mulders,Kreplin2013,Mosoni2013}.
The assumption is that some dynamical process (or a combination of them) lifts dust out of the disk and creates an optically thin halo-like structure, such as magnetocentrifugally driven disk wind \citep{Bans},
buoyancy due to magnetorotational instability \citep{Turner2013},
charged dust in the magnetic field \citep{Ke},
non-radial radiation pressure \citep{Vinkovic09},
and temperature decoupling in the disk surface between dust and gas \citep{Thi}.

Recent near-infrared observations of T Tau stars by \citet{McClure2013a} suggest that the inner rim of disks around their sample of T Tau stars has \hbox{$z_{\rm max}/R_{\rm in}\sim0.2-0.4$} - too high to be explained solely by the puffed-up disk. They used a black-body fit to the near-infrared flux, which is equivalent to assuming that the emission from inner disk wall is dominated by big grains. \citet{VIJE} showed with exact 2D radiative transfer calculations that changing the grain size in the inner disk wall cannot reduce $z_{\rm max}/R_{\rm in}$ and keeps the flux at the same level. However, \citet{McClure2013b} achieved this by using multiple rim walls of different grain size composition. Including viscous heating can explain their success for stars with $\dot{M}_{\rm acc}\gtrsim 10^{-8}M_\odot/$yr, but it is unclear how that worked for lower accretion rates or extreme wall heights. The discrepancy is probably due to their approximation of the inner rim surface as a mixture of grain sizes, while in reality grains separate by size and do not form multigrain surfaces \citep{Vinkovic12}. This approximation overestimates the inner disk radius and boosts the flux. Thus, our work, combined with previous 2D radiative transfer calculations, suggests that the extreme near-infrared excess detected in T Tau stars by \citet{McClure2013a} is difficult to explain by a puffed-up inner rim or its curvature.

\section{Conclusions}

We investigated the inner disk rim height using a self-consistent dust dynamics model based on a long list of physical processes: big dust grains populating the disk surface as the dust grains most resilient to the direct stellar heating, the temperature inversion effect that affects the position of the inner disk radius, accretion luminosity, viscous heating, turbulent diffusion, dust settling due to the gas drag force, vertical hydrostatic equilibrium, radiation pressure force, and the gas velocity profile.
The last two processes have never been included in such an analysis before. The objects under consideration were Herbig Ae stars, T Tau stars, and brown dwarfs.

A wide range of parameters was explored, resulting in more than 6,000 models of the rim height. The results were compared with the rim height implied by the observations of near-infrared excess from protoplanetary disks. Comparison showed that although we built a comprehensive self-consistent model, the modeled rim heights are insufficient to explain the observed flux of Herbig Ae stars. This confirms some previous studies that reached the same conclusion using models that incorporate a smaller list of physical processes.
It also implies that some additional physical processes are influencing dust dynamics in such a way that they increase the overall volume of observed hot optically thin dust.
This problem is closely related to the  question of the shape (i.e. curvature) of the inner disk rim. We showed that radiation pressure creates a cut-off height above which the dust flows outward. This means that the dusty disk cannot exist above this height unless some other forces counterbalance the radiation pressure force. We argued that recent observations of extreme near-infrared excess in T Tau stars are also indicative of extensions of the puffed-up inner rim model.

\begin{acknowledgements}
      The author acknowledges the Technology Innovation Centre Me\dj{}imurje for computational resources. The author thanks the anonymous referee for the critical reviewing of the manuscript and constructive comments.
\end{acknowledgements}

\newpage

\appendix

\section{Theoretical model}
\label{AppendixA}

First we devised a method for calculating the inner disk radius. The traditional approach was to use the dust sublimation temperature $T_{\rm sub}$ as a unique well-defined boundary condition, based on the assumption that the temperature drops with the distance from the star. However, it has been shown numerically \citep{KMD09, Vinkovic12} and theoretically \citep{Vinkovic06} that the temperature of the big grains that populate the surface of the inner disk actually increases with distance within the optically thin dusty zone (see sketch in Figure \ref{FigSketch}) before it starts to monotonically decrease within the optically thick dust. This {\it temperature inversion} effect results in a local temperature maximum within the dust cloud \citep[see Figure 3 in][]{KMD09}, where dust can overheat. A precise calculation of the dust distribution that includes this effect requires information on the spatial dust distribution within the optically thin zone. Since the density structure of the optically thin dust zone cannot be geometrically constrained, we are left with a highly ambiguous concept of the inner disk rim.

\citet{Vinkovic06} proposed using two inner disk radii that distinguish between two coexisting - optically thin and thick - disk zones around a star of luminosity $L_*$:
 \eq{\label{A:Rin_thin_thick}
   R_{\rm in}=0.0344\Psi\left( \frac{1500 K}{T_{\rm sub}} \right)^2 \sqrt{\frac{L_*+L_{\rm acc}}{L_\odot}} \,\,\,\,\,\mathrm{[AU]},
 }
where $R_{\rm in}$ is the disk radius in the midplane, $L_{\rm acc}$ is the accretion luminosity described below and $\Psi=2$ and $\Psi\sim 1.2$ are used for the optically thick and thin inner radii, respectively. Here we considered only optically thick disks with $T_{\rm sub}$ as the dust sublimation temperature of the inner disk rim surface. The sublimation temperature depends on the gas density $\rho_{\rm gas}$. Based on the fit to the data from \citet{Pollack}, we used:
 \eq{\label{A:Tsub}
 T_{\rm sub}=2127\left(\frac{\rho_{\rm gas}}{\mathrm{g/cm^3}}\right)^{0.0216} \mathrm{[K]} .
 }
This is sufficient for the range of gas densities used in our model. A more detailed description of $T_{\rm sub}$ is given by \citet{KMD09}.

The surface of the inner disk is directly exposed to stellar heating, which creates a vertical temperature structure very close to isothermal in the narrow spatial region of our interest. The high-resolution radiative transfer calculation by \citet{Vinkovic12} showed that this is a good approximation for the inner rim of the disk. This approximation can only overestimate the dusty disk height because in reality the temperature slightly drops with the height above midplane. The radial temperature change is approximated with
 \eq{
 T(r)=T_{\rm sub}\left(\frac{r}{R_{\rm in}}\right)^q .
 }

The gas is in vertical hydrostatic equilibrium when the gas density structure is described as
 \eq{\label{A:gas_density}
 \rho_{\rm gas}(r,z)  =  \rho_0 \left(\frac{r}{R_{\rm in}}\right)^p \exp\left(-\frac{z^2}{2H_p^2}\right) ,
 }
 \eq{\label{A:Hp}
 H_p(r)  =  \frac{c_s}{\Omega_K} = 2\times 10^{-3}\sqrt{T_{\rm sub}\frac{M_\odot}{M_*}}*
                                  \left( \frac{r}{R_{\rm in}} \right)^{q/2} \left( \frac{r}{\mathrm{AU}} \right)^{3/2} [\mathrm{AU}] ,
 }
 \eq{\label{A:cs}
 c_s (r) = \sqrt{\frac{k \,\, T(r)}{\mu_{\rm gas}\,m_p}} = 0.0126 \sqrt{T_{\rm sub}} \left( \frac{r}{R_{\rm in}} \right)^{q/2}
                                  \left[\mathrm{ \frac{AU}{year} }\right] ,
 }
 \eq{\label{A:OmegaK}
 \Omega_K\,(r) = 2\pi \sqrt{\frac{M_*}{M_\odot}} \left( \frac{\mathrm{AU}}{r}\right)^{3/2} \left[\mathrm{\frac{1}{year} }\right],
 }
where $\rho_0 = \rho_{\rm gas}(R_{\rm in},0)$ is the midplane gas density at the inner disk rim, $H_p$ is the disk scale height, $c_s$ is the isothermal sound speed, $\Omega_K$ is the Keplerian angular velocity at the midplane, $M_*$ is the stellar mass, $\mu_{\rm gas}=2.33$ is the mean molecular weight in units of the proton mass $m_p$, and $k$ is Boltzmann's constant.

\citet{TakeuchiLin02} showed that the vertical component of the gas velocity is much smaller than the horizontal component $\mathrm{v}_{\rm r,gas}$, given by
 \eq{\label{A:V_r}
 \mathrm{v}_{\rm r,gas}(r,z)=- \frac{\alpha c_s H_p}{r}\left( 3p+2q+6+\frac{5q+9}{2}\frac{z^2}{H_p^2} \right)
                   \left[\mathrm{ \frac{AU}{year} }\right].
 }

Accretion of the gas onto the star contributes with its luminosity to the total energy that heats the inner rim. To accommodate for this effect we used the accretion luminosity \citep{Calvet,Muzerolle} in equation \ref{A:Rin_thin_thick}
 \eq{\label{A:Lacc}
 L_{\rm acc}=\xi\frac{GM_*\dot{M}_{\rm acc}}{R_*}=3.18 \xi \,\left(\frac{M_*}{M_\odot }\right)\left(\frac{R_\odot}{R_*}\right)
          \left(\frac{\dot{M}_{\rm acc}}{10^{-7}M_\odot/year}\right) [L_\odot] ,
 }
where $M_*$ is the stellar mass, $\xi\sim 0.8$ is the correction factor for the magnetospheric truncation of the disk, $G$ is the gravitational constant, and $R_*$ is the stellar radius.

The accretion rate $\dot{M}_{\rm acc}$ can be calculated from the vertical profile of the gas density (eq.\ref{A:gas_density}) and the horizontal velocity (eq.\ref{A:V_r})
 \eq{\label{A:Macc}
 \dot{M}_{\rm acc}(r) = 2 \int_0^\infty \rho_{\rm gas}(r,z)\,\mathrm{v}_{\rm r,gas}(r,z)\, 2\pi\, r\, dz ,
 }
which exists as inflow only if
 \eq{
 p+\frac{3}{2}(q+1) > -2 .
 }

The accretion rate depends on the distance from the star, which drives the disk evolution, except for $q=-0.5$ when the radial velocity is independent of $r$ and the disk is in steady-state. We focused our attention on a very small part of the disk around $R_{\rm in}$, where $\dot{M}_{\rm acc}$ varies slowly. Hence, we define the accretion rate as $\dot{M}_{\rm acc}(R_{\rm in})$. Equation \ref{A:Macc} also shows that $\dot{M}_{\rm acc}\propto \rho_0$. We used this to derive $\rho_0$ from $\dot{M}_{\rm acc}$, which is in turn derived from $L_{\rm acc}\propto \dot{M}_{\rm acc}$ in equation \ref{A:Lacc}.

The local viscous dissipation of energy in the accretion disks contributes to the local disk temperature, and we checked for this effect. The net temperature at $R_{\rm in}$ is the combination $T_{\rm sub}^4=T_{\rm irr}^4+T_{\rm visc}^4$  \citep{CieslaCuzzi} of the stellar irradiation (equation \ref{A:Rin_thin_thick}, with $T_{\rm sub}$ replaced by $T_{\rm irr}$) and viscous heating
 \eq{\label{A:Tvisc}
 T_{\rm visc}^4 = \frac{3GM_*\dot{M}_{\rm acc}}{8\pi\sigma r^3}=151^4\left(\frac{M_*}{M_\odot}\right)
              \left(\frac{\dot{M}_{\rm acc}}{10^{-7}M_\odot/year}\right)\left(\frac{\mathrm{AU}}{R_{\rm in}}\right)^3 ,
 }
where $\sigma$ is the Stefan-Boltzmann constant. If $T_{\rm visc}^4\ll T_{\rm sub}^4$, we used equation \ref{A:Rin_thin_thick} to derive $R_{\rm in}$ and if $T_{\rm visc}^4\gg T_{\rm sub}^4$, we used equation \ref{A:Tvisc}. These conditions have to be checked iteratively when calculating $R_{\rm in}$.

The long-term stability of the inner disk rim depends on the radial migration of the dust particles. We followed the prescription by \citet{TakeuchiLin02} where dust particles are driven by gas drag and their radial velocity depends on the distance from the midplane. The source of this vertical velocity variation is the gas pressure gradient, which causes two effects: the gas rotation is different from Keplerian rotation above the midplane, and the gas drag force depends on gas density.
Without the gas drag the dust particles would follow Keplerian orbits. We considered particles smaller than the mean free path of gas molecules. This condition is described with Epstein's gas drag law that gives the gas drag stopping time:
 \eq{\label{A:tdrag}
 t_{\rm drag}(r,z) = \frac{\rho_{\rm grain} \,\,a}{\rho_{\rm gas} v_{\rm th}} = t_0(r)  \exp\left(\frac{z^2}{2H_p^2}\right) ,
 }
 \eq{\label{A:tdrag0}
 t_0 (r) = \frac{10^{-15}}{3}\left(\frac{\rho_{\rm grain}}{\rho_0}\right)
      \left(\frac{a/\mu m}{\sqrt{T_{\rm sub}}}\right) \left(\frac{R_{\rm in}}{r}\right)^{p+q/2}  [\mathrm{year}] ,
 }
where $\rho_{\rm grain}$ is the dust grain solid density, $a$ is the dust grain radius, and $v_{\rm th}=\sqrt{8/\pi}\,\,c_s$ is the mean thermal velocity of gas.

We considered turbulent disks where the gas turbulent motion stirs up dust particles to higher altitudes to prevent dust sedimentation. The equilibrium distribution of dust particles is reached when the sedimentation due to gas drag and the diffusion due to turbulence are balanced. Using the $\alpha$-prescription for the viscous effect of turbulence \citep{ShakuraSunyaev,Pringle}, the dust density distribution is \citep{TakeuchiLin02}
 \eq{
 \rho_{\rm dust}(r,z)=\rho_d \exp\left( -\frac{z^2}{2H_p^2} - \frac{S_c\,\Omega_K\,\, t_0(r)}{\alpha}
                  \left(\exp\frac{z^2}{2H_p^2}-1\right) \right),
 }
where $\rho_d(r)$ is the midplane dust density profile. The Schmidt number $S_c$ is a measure of the coupling between the particles and the gas. $S_c\sim 1$ for the small particles used in our paper and increases to infinity for very big particles. This means that very big particles are not influenced by the gas turbulence and they accumulate in the midplane. The expression of $S_c$ is a matter of debate, and we adopted the expression suggested by \citet{Youdin} for diffusion driven by anisotropic turbulence
 \eq{\label{A:Sc}
 S_c = \frac{\left( 1+\Omega_K^2 \,t_{\rm drag}^2 \right)^2}{1+4\Omega_K\, t_{\rm drag}}.
 }

\citet{TakeuchiLin03} showed that the stellar radiation pressure force slowly erodes the dusty disk surface, but they did not investigate the impact of this force on the surface of inner disk rim. We used their methodology to include the radiation pressure force in the dust velocity profiles. We ignored the complicated radiation pressure force from the near-infrared photons from the disk interior \citep{Vinkovic09}, but this force can only enhance the disk erosion and reduce the height of the optically thick disk. The stellar radiation pressure force compares to the stellar gravity by a factor of \citep{Vinkovic09}
 \eq{\label{A:beta}
 \beta = 0.6 \left(\frac{L_*+L_{\rm acc}}{L_\odot}\right) \left(\frac{M_\odot}{M_*}\right)
             \left(\frac{\mathrm{g/cm^3}}{\rho_{\rm grain}}\right) \frac{\langle Q^{\rm ext}\rangle}{\mathrm{a/\mu m}} ,
 }
where $\langle Q^{\rm ext}\rangle$ is the dust extinction coefficient averaged over the stellar spectrum. We used a stellar black-body spectrum to calculate $\langle Q^{\rm ext}\rangle$ from olivine dust grains. However, the result is close to $\langle Q^{\rm ext}\rangle\sim 2$ because the stellar radiation peaks at wavelengths shorter than the size of big grains that populate the inner disk surface. The radial component of the dust velocity with included radiation pressure was derived by \cite{TakeuchiLin03} as
 \eq{\label{A:v_r_dust}
 \mathrm{v}_{\rm r,dust}(r,z)=\frac{\mathrm{v}_{\rm r,gas}+(\beta-\eta)\,r\,t_{\rm drag}\,\Omega_K^2}
                               {1+t_{\rm drag}^2\Omega_K^2} \left[\mathrm{ \frac{AU}{year} }\right] ,
 }
 \eq{
 \eta = -\left(p+q+\frac{q+3}{2}\frac{z^2}{H_p^2}\right) \left(\frac{H_p}{r}\right)^2 .
 }

Our goal is to find the relative height $z_{\rm max}/R_{\rm in}$ that separates the inflow below this height from the outflow \hbox{$\mathrm{v}_{\rm r,dust}(R_{\rm in},z_{\rm max})>0$} above $z_{\rm max}$. We varied the parameters listed in Table \ref{parameters} and iteratively searched for $\rho_0$ consistent with $\dot{M}_{\rm acc}$, which in turn has to be consistent with $L_{\rm acc}$. The pseudocode describing the computational procedure is the following:

{\small
  \begin{algorithmic}[1]
  \STATE Set initial parameters
    \FOR {loop over grain size $a$}
      \STATE get $\langle Q^{\rm ext}\rangle$ for this grain
      \FOR {loop over $q$}
      \FOR {loop over $p$}
      \IF {$p+\frac{3}{2}(q+1) > -2$}
      \FOR {loop over $\rho_{\rm grain}$}
      \FOR {loop over $\alpha$}
      \FOR {loop over $L_{\rm acc}$}
      \STATE set $\rho_0$ to some very low initial value
      \REPEAT
      \STATE find $T_{\rm sub}$ from eq.\ref{A:Tsub}
      \STATE find $R_{\rm in}$ from eq.\ref{A:Rin_thin_thick}
      \STATE find $H_p$ from eq.\ref{A:Hp}
      \STATE find $c_s$ from eq.\ref{A:cs}
      \STATE find $\Omega_K$ from eq.\ref{A:OmegaK}
      \STATE find $t_0$ from eq.\ref{A:tdrag0}
      \STATE find $\beta$ from eq.\ref{A:beta}
      \STATE set $\dot{M}_{\rm acc}=0$
      \STATE set step height $\Delta (z/H_p)$ for integration
      \FOR {loop over $(z/H_p)$ from 0 to
      some upper integration limit}
        \STATE find $t_{\rm drag}(z)$ from eq.\ref{A:tdrag}
        \STATE find $S_c(z)$ from eq.\ref{A:Sc}
        \STATE find $\mathrm{v}_{\rm r,gas}(z)$ from eq.\ref{A:V_r}
        \STATE find $\mathrm{v}_{\rm r,dust}(z)$ from eq.\ref{A:v_r_dust}
        \IF {$\mathrm{v}_{\rm r,dust}(z)$ switched to outflow}
          \STATE memorize $z_{\rm max}\equiv z$
        \ENDIF
        \STATE increase $\dot{M}_{\rm acc}$ using eq.\ref{A:Macc}
      \ENDFOR
      \STATE find $L_{\rm acc}^{\rm ref}$ from eq.\ref{A:Lacc}
      \STATE check for viscous heating using eq.\ref{A:Tvisc}
      \STATE adjust $\rho_0$ as necessary for $L_{\rm acc}^{\rm ref}\rightarrow L_{\rm acc}$
      \UNTIL {$L_{\rm acc} == L_{\rm acc}^{\rm ref}$ OR
      $\rho_0 \ge $ some upper limit }
      \STATE print $z_{\rm max}/R_{\rm in}$
      \ENDFOR
      \ENDFOR
      \ENDFOR
      \ENDIF
      \ENDFOR
      \ENDFOR
    \ENDFOR
    \medskip
  \end{algorithmic}
}

\end{document}